\documentclass[preprint,aps]{revtex4}
\usepackage{graphicx}
\usepackage{dcolumn}
\usepackage{bm}
\usepackage{xcolor}
\begin{document}

\title{Superconductivity in ThMo$_{2}$Si$_{2}$C with Mo$_2$C Square Net}{Superconductivity in ThMo$_{2}$Si$_{2}$C with Mo$_2$C Square Net}

\author{Zichen Liu,$^{1}$ Baizhuo Li,$^{2}$ Yusen Xiao,$^{1}$ Qingchen Duan,$^{1}$ Yanwei Cui,$^{3}$ YuXue Mei$^{1}$ Qian Tao,$^{2}$ Shuli Wei,$^{1}$ Shugang Tan,$^{1}$ Qiang Jing,$^{1}$ Qing Lu,$^{1}$ Yuping Sun,$^{1}$ Yunyan Liu,$^{1}$ Shenggui Fu,$^{1}$ Hao Jiang,$^{4}$ Zhi Ren,$^{3}$  Zhu'an Xu,$^{2}$ Cao Wang,$^{1}$\footnote{E-mail: wangcao@sdut.edu.cn} Guanghan Cao.$^{2}$
}

\address{$^{1}$School of Physics and Optoelectronic Engineering, Shandong University of Technology, Zibo 255000, P. R. China\\
         $^{2}$Department of Physics, Zhejiang Province Key Laboratory of Quantum Technology and Devices, Interdisciplinary Center for Quantum Information, and State Key Lab of Silicon Materials, Zhejiang University, Hangzhou 310027, P. R. China\\
         $^{3}$School of Science, Westlake University, 18 Shilongshan Road, Hangzhou 310064, P. R. China\\
         $^{4}$School of Physics and Optoelectronics, Xiangtan University, Xiangtan 411105, P. R. China}

\date{\today}

\begin{abstract}
We report the superconductivity of a new quaternary compound ThMo$_2$Si$_2$C, synthesized with the arc-melting technique. The compound crystallizes in a tetragonal CeCr$_2$Si$_2$C-type structure with cell parameters of $a$ = 4.2296 {\AA} and $c$ = 5.3571 {\AA}. An interlayer Si--Si covalent bonding is suggested by the atomic distance. The electrical resistivity and magnetic susceptibility measurements indicate a Pauli-paramagnetic metal with dominant electron--electron scattering in the normal-state. Bulk superconductivity at 2.2 K is demonstrated with a dimensionless specific-heat jump of ${\Delta}C/\gamma_{\rm n}T$ = 0.98. The superconducting parameters of the critical magnetic fields, coherence length, penetration depth, and superconducting energy gap are given.


\end{abstract}



\maketitle
\section{\label{sec:level1}Introduction}

Since the discovery of high-temperature cuprate superconductors, the exploration of new superconductors has aroused enormous interest in layered compounds, including $R$Ni$_2$B$_2$C($R$ for rare earth),\cite{RNi2B2C} MgB$_2$,\cite{MgB2} Sr$_2$RuO$_4$,\cite{Sr2RuO4} iron pnictides/chalcogenides,\cite{LaFeAsO, FeSe} and BiS$_2$-based compounds.\cite{BiS2} Recently, a new quaternary superconductor BaTi$_2$Sb$_2$O has been reported, consisting of alternately stacked Ba planes and Ti$_2$Sb$_2$O layers. Upon cooling, the compound undergoes a density-wave-like (DW) transition at 54 K and superconducting transition at 1.2 K.\cite{BaTi2Sb2O, BaTi2Sb2O-NMR,BaTi2Sb2O-muSR, BaTi2Sb2O-FDL} The DW ordering can be suppressed by both hole doping in Ba$_{1-x}$Na$_x$Ti$_2$Sb$_2$O and isovalent substitution in BaTi$_2$Sb$_{2-x}$Bi$_x$O, where the optimal superconducting $T_{\rm c}$s reaches 5.5 K and 3.7 K, respectively.\cite{BaTi2Sb2O-Na,BaTi2Sb2O-Bi} Superconductivity of 4.6 K has also been observed in the undoped end member BaTi$_2$Bi$_2$O.\cite{BaTi2Bi2O} All these discoveries mark the birth of a new superconducting family based on the Ti$_2Pn_2$O ($Pn$ = Sb, Bi) block layers.\cite{Ti2Pn2O-review} If the single-layered Ba atom in the lattice is replaced by fluorite-like Sr$_2$F$_2$ or Sm$_2$O$_2$ layers, no superconductivity is observed while the DW state prevails.\cite{BaTi2Bi2O,Sm2O2Ti2Sb2O} In addition, the coexistence of superconductivity and DW transition was observed in the intergrowth compound Ba$_2$Ti$_2$Fe$_2$As$_4$O, where the superconductivity and the DW instability were ascribed to the Fe$_2$As$_2$ and Ti$_2$As$_2$O layers, respectively.\cite{SunYL}

The newly discovered superconductors containing Ti$_2Pn_2$O layers constitute a small fraction of a much larger compound family. Among them, the key structural motif is the anti-CuO$_2$-layer-like $M_2$O square planes, such as Ti$_2$O,\cite{Ti2Pn2O-review} V$_2$O,\cite{CsV2S2O, RbV2Te2O} Mn$_2$O,\cite{La2O3Mn2Se2} Fe$_2$O,\cite{La2Fe2Se2O3, Sr2F2Fe2Se2O, Na2Fe2Se2O, BaFe2Se2O, Ca2O3Fe3S2} and Co$_2$O.\cite{La2Co2Se2O3} The $M_2$O planes are sandwiched by the $X$ (chalcogen, pnictogen) monolayers, forming unique $M_2X_2$O structural units. We note that a similar $M_2X_2$O-type structure can also be found in oxygen-free compounds, such as LaRu$_2$Al$_2$B and $RM_2$Si$_2$C ($M$ = Cr, Mo).\cite{LaRu2Al2B, CeCr2Si2C, RCr2Si2C, CeMo2Si2C,PrMo2Si2C} The most distinctive feature of these oxygen-free compounds, compared to oxygen-containing ones, is the $X$--$X$ interlayer covalent bond, which results in lattice collapse along the $c-$axis. All these compounds with a collapsed structure show metallic behavior, and the electronic states observed in the vicinity of the Fermi level are mainly dominated by $d-$orbital electrons.\cite{LaRu2Al2B,CeMo2Si2C,RCr2Si2C-PRB} To date, no superconductivity has been reported in these compounds.

Here, we report the synthesis, structure, and physical properties of a new layered carbide ThMo$_2$Si$_2$C. Powder X-ray diffraction (XRD) indicates that the compound crystallizes in the CeCr$_2$Si$_2$C-type structure with interlayer Si--Si covalent bonds. The compound exhibits metallic behavior with Pauli paramagnetism. Both magnetic susceptibility and electronic resistivity reveal superconductivity of $T_{\rm c}$ = 2.2 K. No charge/spin-density-wave anomaly is observed on the $\chi-T$ and $\rho-T$ curves, which marks the difference to BaTi$_2$Sb$_2$O. The compound is characterized as a phonon-mediated superconductor with considerable electron--electron interaction.

\section{\label{sec:level2}Experimental}

\begin{figure}[h]
\centering
\includegraphics[scale=0.7]{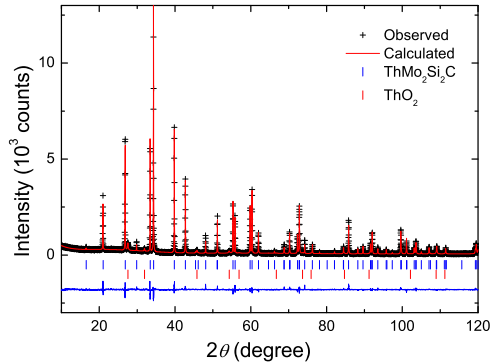}
\caption{(Color online) Rietveld refinement profile of ThMo$_2$Si$_2$C polycrystalline sample.}
\label{fig1}
\end{figure}

A polycrystalline sample of ThMo$_2$Si$_2$C was synthesized using powders of Th, Mo(Alfa 99.9\%), Si (Alfa 99.99\%), and graphite (99.95\%) as starting materials. The preparation of thorium metal is described elsewhere.\cite{ThFeAsN} The stoichiometric mixture of the materials were sufficiently mixed in an agate mortar and cold-pressed into a pellet. All these procedures are carried out in a glove box protected by high purity argon. The pellet was then melted by the standard arc-melting technique on a water-cooled copper hearth. The sample was melted five times and was flipped after each melting process to ensure homogeneity. The arc-melted button was placed in an alumina crucible, sealed in an evacuated silica ampoule, and annealed at 1100 $^{\circ}$C for two weeks.

Powder XRD was carried out at room temperature on a PANalytical X-ray diffractometer (model EMPYREAN) with monochromatic Cu K$\alpha1$ radiation. The crystal structure data were obtained by Rietveld refinement using the step-scan XRD data with $10^{\circ}\leq2\theta\leq120^{\circ}$, and the diffraction data were refined via the Rietveld method using the program \textit{RIETAN-2000}.\cite{RIETAN-2000} The structural parameters were obtained assuming that the occupancy was the same as the chemical composition. The electrical resistivity was measured in a Quantum Design Physical Property Measurement System (PPMS-9 Dynacool), using a standard four-probe method down to 1.8 K, and then, if necessary, from 1.8 to 0.4 K with an adiabatic dilution refrigerator option. The measurement of specific heat down to 0.4 K with $^3$He subassembly was carried out on a Quantum Design PPMS-9 Evercool II using a relaxation method. The dc magnetization measurements down to 0.4 K were performed on a Quantum Design magnetic property measurement system (MPMS3) with a $^3$He cryostat.

\section{\label{sec:level3}Results and discussion}

\begin{table}[h]
\footnotesize
\caption{Summary of crystallographic data and refinement parameters for ThMo$_2$Si$_2$C}\label{tab1}
\doublerulesep 0.1pt \tabcolsep 14pt
\begin{tabular}{cccccc}

  & Data set       &&& Parameters        &\\
  \hline
  &space group     &&& $P4/mmm$          & \\
  &$a$ (\AA)     &&& 4.2296(3)      & \\
  &$c$ (\AA)     &&& 5.3571(4)       & \\
  &$V$ (\AA$^3$) &&& 95.84(1)        & \\
  &$R_{\rm{wp}}$ (\%)   &&& 8.346          & \\
  &$S$             &&& 1.21            & \\
  &$Z$             &&& 1               & \\
  \hline

\end{tabular}
\end{table}

\begin{figure}[h]
\centering
\includegraphics[scale=0.7]{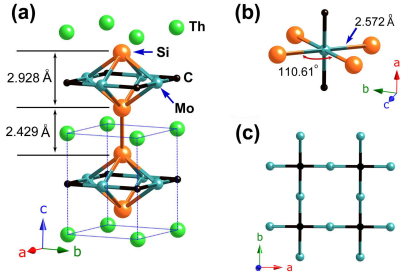}
\caption{(Color online) (a) Crystal structure of ThMo$_2$Si$_2$C. (b) Schematic of the distorted MoSi$_4$C$_2$ octahedron. (c) Bird's-eye view of the Mo$_2$C square lattice.}
\label{fig2}
\end{figure}

Figure \ref{fig1} shows the XRD pattern of ThMo$_2$Si$_2$C together with the Rietveld refinement profile for the CeCr$_2$Si$_2$C-type structure (space group $P4/mmm$). The obtained lattice parameters for ThMo$_2$Si$_2$C are $a$ = 4.2296(3) {\AA} and $c$ = 5.3571(4) {\AA}. A small amount of ThO$_2$ (3.7 wt.\%, according to refinement) is present, which can be ascribed to oxygen contamination during the sample transfer from the glove box to the arc-melting equipment. Table \ref{tab1} lists the refined crystal structure parameters. Table \ref{tab2} shows the coordinates and temperature factors for all the atoms in the lattice. Figure \ref{fig2} shows the crystal lattice of ThMo$_2$Si$_2$C and the details of the Mo$_2$Si$_2$C block layer. Table \ref{tab3} compares the lattice parameters of ThMo$_2$Si$_2$C with those of the isostructural compounds CeMo$_2$Si$_2$C and PrMo$_2$Si$_2$C. According to the refinement, the interlayer's nearest Si--Si distance ($d_{\rm{Si-Si}}$) is 2.429 {\AA}. This Si--Si distance is similar to that of silicon with a diamond-type structure(2.352 {\AA}), and thus should be considered a single Si--Si bond. Therefore, the structure of ThMo$_2$Si$_2$C is of the CeCr$_2$Si$_2$C-type collapsed phase. As is well known, the ionic radius of Th$^{4+}$ (0.94 {\AA}, $CN$ = 6) is smaller than that of Pr$^{3+}$ (0.99 {\AA}) or Ce$^{3+}$ (1.01 {\AA}).\cite{radii} However, we note that the cell volume does not follow the radius trend of the rare-earth elements. Moreover, the $a-$axis of ThMo$_2$Si$_2$C is the longest among the three compounds. This can be understood as an electron-doping effect, as Th$^{4+}$ contributes more electrons than Pr$^{3+}$ or Ce$^{3+}$. On the one hand, the electron transferred to the Mo$_2$C plane expands the $a-$axis by elongating the Mo--C bond. On the other hand, the charge transfer strengthens the interlayer Coulomb attraction and results in the shrinkage of the $c-$axis. We note that similar phenomena are quite common in iron arsenide superconductors, where electron doping can cause a decrease in axial ratio $c/a$.\cite{Gd1111-Th, La1111-Co}

\begin{table}[h]
\footnotesize
\caption{Atomic coordinates and temperature factors for all the elements in ThMo$_2$Si$_2$C}\label{tab2}
\doublerulesep 0.1pt \tabcolsep 11pt
\begin{tabular}{cccccc}

  atom  &  site    &  $x$  &  $y$  &  $z$        &  $B$          \\
  \hline
  Th    &  1a      &  0    &   0   &   0         &  0.1(fixed) \\
  Mo    &  2e      &  0    &   1/2 &   1/2       &  0.15(3)    \\
  Si    &  2h      &  1/2  &   1/2 &   0.2267(6) &  0.17(7)    \\
  C     &  1b      &  0    &   0   &   1/2       &  1.9(5)     \\
  \hline
\end{tabular}

\end{table}

\begin{table*}[t]
\footnotesize
\caption{Lattice parameters of $R$Mo$_2$Si$_2$C ($R$ = Ce, Pr, Th), where $h_{\rm Si}$ and $d_{\rm{Si-Si}}$ represent the vertical distance from Si to the Mo$_2$C plane and the interlayer Si--Si distance.}
\label{tab3}
\tabcolsep 12pt
\begin{tabular*}{\textwidth}{lcccccc}
Compound                           &  $a-$axis ({\AA})    &  $c-$axis ({\AA})  &  $c/a$   &   $V_{\rm{cell}}$ ({\AA}$^3$)  &  $h_{\rm Si}$ ({\AA})  & $d_{\rm{Si-Si}}$ ({\AA})  \\
  \hline
  CeMo$_2$Si$_2$C\cite{CeMo2Si2C}    &  4.213          &  5.376        &   1.276  &   95.42            &  1.512             &   2.353              \\
  PrMo$_2$Si$_2$C\cite{PrMo2Si2C}    &  4.2139         &  5.4093       &   1.284  &   96.05            &  1.508             &   2.393              \\
  ThMo$_2$Si$_2$C                    &  4.2296         &  5.3571       &   1.267  &   95.84            &  1.464             &   2.429              \\
  \hline
\end{tabular*}
\end{table*}

\begin{figure}[h]
\centering
\includegraphics[scale=0.7]{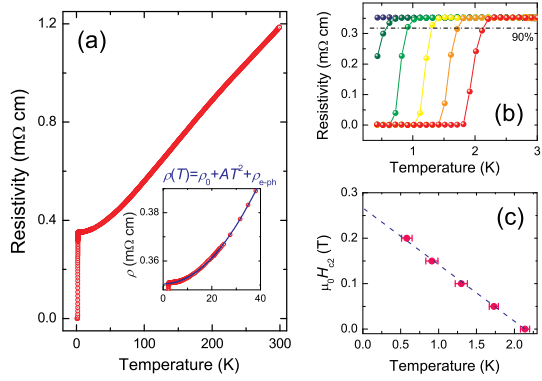}
\caption{(Color online) (a) Temperature dependence of resistivity at zero magnetic field for the ThMo$_2$Si$_2$C polycrystalline sample. The inset shows the normal-state resistivity below 40 K. The blue line fits the data using the equation $\rho(T)=\rho_0+AT^2+\rho_{\rm e-ph}$, where $\rho_{\rm e-ph}$ represents the contribution of electron--phonon scattering. (b) Temperature dependence of resistivity near the superconducting transition measured in selected magnetic fields. (c) Temperature dependence of the upper critical field.}
\label{fig3}
\end{figure}

Figure \ref{fig3} shows the resistivity data of the ThMo$_2$Si$_2$C sample. The zero-field resistivity at 300 K is 1.19 m$\Omega\cdot$cm, which is similar to that of CeMo$_2$Si$_2$C.\cite{CeMo2Si2C} Upon cooling, the resistivity exhibits metallic behavior without anomaly. A superconducting transition is revealed by an abrupt resistivity drop at 2.2 K. The residual resistivity is high, suggesting that the sample may be in the dirty limit. The normal-state resistivity below 40 K can be fitted using the equation
\begin{equation}
\rho(T)=\rho_0+AT^2+\alpha \left ( \frac{T}{\Theta_R} \right )^n \int^{\Theta_R/T}_{0}\frac{x^n}{(e^x-1)(1-e^{-x})}dx
\end{equation}
where $\Theta_R$ is the Debye temperature obtained from the resistivity measurements and matches very closely with that obtained from the specific heat. The best fit appears while the constant $n$ in the last term of the equation (contribution of electron--phonon scattering) is 2. The fitting parameters are $\rho_0$ = 0.35 m$\Omega\cdot$cm and $A$ = 2.75$\times10^{-5}$ $m\Omega\cdot\rm{cm}\cdot K^{-2}$. Figure \ref{fig3}(b) shows the temperature dependence of resistivity under static magnetic fields (0 T, 0.05 T, 0.1 T, 0.15 T, 0.2 T, 0.25 T). We define $T_{\rm c}(H)$ as the temperature where the resistivity falls to 90\% of the normal-state value. As shown in Figure \ref{fig3}(c), the temperature dependence of the upper critical field ($H_{\rm c2}$) increases approximately linearly upon cooling to the lowest temperature measured. Thus, the upper critical field at 0 K is estimated from linear extrapolation to be $H_{\rm c2}(0)$ = 2660 Oe, and the coherent length can be calculated as $\xi$ = 352 {\AA} according to the relation $\mu_0H_{\rm c2}(0) = \Phi_0/2\pi\xi^2$, where $\Phi_0$ is the magnetic flux quantum.

\begin{figure}[h]
\centering
\includegraphics[scale=0.7]{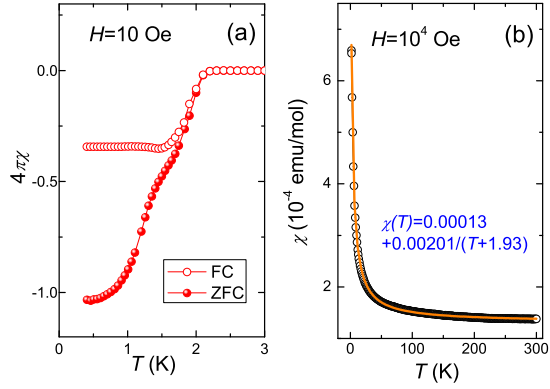}
\caption{(Color online) (a) Temperature dependence of magnetic susceptibility measured at $H$ = 10 Oe for ThMo$_2$Si$_2$C. (b) Magnetic susceptibility measured at $H$ = 10$^4$ Oe. The orange solid line fits the normal-state $\chi-T$ curve using the equation $\chi(T) = \chi_0+C/(T-\theta)$.}
\label{fig4}
\end{figure}

Figure \ref{fig4}(a) plots the temperature dependence of magnetic susceptibility at $H$ = 10 Oe. Both zero-field-cooling (ZFC) and field-cooling (FC) protocols are employed in the measurement. A strong diamagnetic signal below 2.2 K confirms the superconducting transition observed in the resistivity measurement. The magnetic shielding fraction at 0.4 K exceeds 100\% due to the demagnetization effect, indicating bulk superconductivity. The shoulder-like susceptibility (ZFC) around 1.5 K is probably due to the intergrain weak-link effect. The susceptibility measured at $H$ = 10$^4$ Oe is shown in Figure \ref{fig4}(b). The room temperature susceptibility is 1.38$\times10^{-4}$ emu$\cdot$mol$^{-1}$. The susceptibility increases gradually upon cooling. We fit the susceptibility data using the extended Curie--Weiss law $\chi = \chi_0+C/(T-\theta)$ over a temperature region of 10 K $\leq T \leq$ 300 K. The derived parameters are $\chi_0$ = $1.3\times10^{-4}$ emu$\cdot$mol$^{-1}$, $C$ = $2.01\times10^{-3}$ emu$\cdot$K$\cdot$mol$^{-1}$, and $\theta$ = $-1.9$ K. The results suggest that ThMo$_2$Si$_2$C is a metal with Pauli paramagnetism. The Curie constant corresponds to a tiny effective magnetic moment of 0.12 $\mu_{\rm B}$/f.u., which may be due to a small amount of paramagnetic impurities.

\begin{figure}[h]
\centering
\includegraphics[scale=0.7]{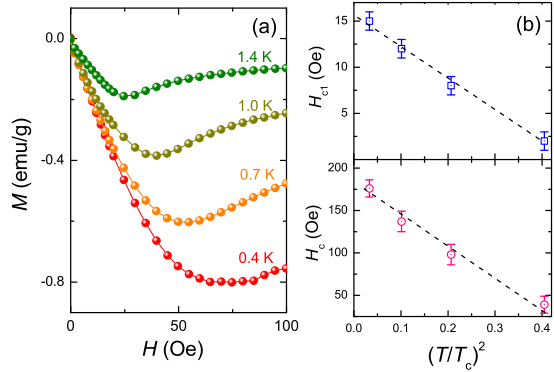}
\caption{(Color online) (a) Field dependence of magnetization for ThMo$_2$Si$_2$C at various temperatures. (b) The lower critical field $H_{\rm c1}$ and the thermodynamic critical field $H_{\rm{c}}$ for ThMo$_2$Si$_2$C as a function of $(T/T_{\rm c})^2$.}
\label{fig5}
\end{figure}

Figure \ref{fig5} shows the field dependence of the zero-field cooled magnetization at various temperatures for ThMo$_2$Si$_2$C. The lower critical field $H_{c1}$ is plotted as a function of ($T/T_{\rm c}$)$^2$. We fit the data using the Ginzburg--Landau theory $H_{\rm c1}(T)$ = $H_{\rm c1}(0)[1-(T/T_{\rm c})^2]$ (where $T_{\rm c}$ = 2.2 K) and obtain the lower critical field $H_{\rm c1}(0)$ = 16 Oe. Then the penetration depth and the Ginzburg--Landau parameter are derived as $\lambda$ = 5280 {\AA} and $\kappa$ = 15 from the relation $\mu_0H_{\rm c1}(0)$ = $ln(\lambda/\xi)\Phi_0/(4\pi\lambda^2)$ and $\kappa$ = $\lambda/\xi$, respectively. Also, the thermodynamic critical field at zero temperature $H_{\rm c}$(0) is estimated as 183 Oe.

\begin{figure}[h]
\centering
\includegraphics[scale=0.7]{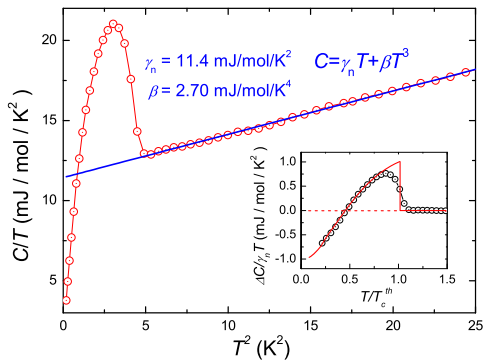}
\caption{(Color online) The specific heat coefficient $C$/$T$ of ThMo$_2$Si$_2$C as a function of $T^2$. The inset plots the normalized electronic specific heat ${\Delta}C/\gamma_{\rm n}T$ after subtracting the normal-state contribution. The solid red line presents the fit with the conventional BCS model.}
\label{fig6}
\end{figure}

Figure \ref{fig6} shows the temperature dependence of the specific heat measured under zero magnetic field. To make the specific heat of normal-state more intuitive, we plot $C/T$ as a function of $T^2$ from 5 K down to 0.4 K. The data show good linear $T^2$ dependence of $C/T$ above superconducting $T_{\rm c}$, indicating that the specific heat comprises two contributions: the electronic part, which is proportional to $T$, and the phonon part, which is proportional to $T^3$ at low temperatures. We fit the specific heat of the normal-state below 5 K using the relation $C = \gamma_{\rm n}T+{\beta}T^3$. The parameters of the electron and phonon are determined as $\gamma_{\rm n}$ = 11.4 mJ${\cdot}$mol$^{-1}{\cdot}$K$^{-2}$ and $\beta$ = 2.7 mJ$\cdot$mol$^{-1}{\cdot}$K$^{-4}$ respectively. According to the formula $\Theta_{\rm D}$ = $[12\pi^4NR/(5\beta)]^{1/3}$, the Debye temperature $\Theta_{\rm D}$ is estimated to be 351 K.

Below the $T_{\rm c}$, a distinct peak in specific heat is observed. We subtract the normal-state specific heat using the aforementioned fitted data $C = \gamma_{\rm n}T+{\beta}T^3$. The inset of Figure \ref{fig6} plots ${\Delta}C/\gamma_{\rm n}T$ versus $T/T_{\rm c}^{\rm th}$, where $T_{\rm c}^{\rm th}$ = 2 K denotes the thermodynamic critical temperature considering entropy conservation. Supposing that an isotropic single-gap model can be adopted here, we fit the specific heat below 1.6 K with the exponential law $C_{\rm e} \sim {\rm exp}(-\Delta_0/k_{\rm B}T)$. The derived energy gap is $\Delta_0^{\rm exp}$ = 0.29 meV, which matches the Bardeen--Cooper--Schrieffer (BCS) value of 0.30 meV based on the relation $\Delta_0^{\rm BCS}$ = $1.76k_{\rm B}T_{\rm c}^{\rm th}$. Further, the electronic specific heat jump at $T_{\rm c}^{\rm th}$ brings ${\Delta}C/\gamma_{\rm n}T$ = 0.98, which is smaller than the BCS value of ${\Delta}C/\gamma_{\rm n}T$ = 1.43. A possible explanation for this deviation is the presence of nonsuperconducting impurities (e.g., ThO$_2$) and/or amorphous grain-boundary that does not contribute to the signal of superconducting transition. Assuming that the electron--phonon coupling dominates the origin of the superconductivity in ThMo$_2$Si$_2$C, the coupling strength $\lambda_{\rm ph}$ can be estimated from the McMillan formula
\begin{equation}
\lambda_{\rm ph} = \frac{1.04+\mu^*\ln(\frac{\Theta_{\rm D}}{1.45T_{\rm c}})}{(1-0.62\mu^*)\ln(\frac{\Theta_{\rm D}}{1.45T_{\rm c}})-1.04}
\end{equation}
where $\mu^*$ = 0.13 is the Coulomb pseudopotential for polyvalent transition metals.\cite{McMillan} Using this formula, $\lambda_{\rm ph}$ was evaluated to be 0.49, suggesting that the superconductivity in ThMo$_2$Si$_2$C is in the weak or intermediate coupling regime.

The Wilson ratio ($R_{\rm W}$) that reflects the electron--electron correlations in a material can be calculated using the formula $R_{\rm W} = \chi_{\rm{P}}/(3\gamma_{\rm n})\times(\pi{k_{\rm B}}/\mu_{\rm B})^2$, where $\chi_{\rm{P}}$ represents the Pauli paramagnetic susceptibility.\cite{wilson} During the Curie-Weiss fitting of the magnetic susceptibility, we get the temperature independent susceptibility $\chi_0$ = $1.3\times10^{-4}$ emu$\cdot$mol$^{-1}$. Note that the fitted $\chi_0$ includes the orbital diamagnetism contribution due to ion cores $\chi_{\rm{core}}$. If the nominal valences can be represented by the formula Th$^{4+}$Mo$^{3+}_2$Si$^{3-}_2$C$^{4-}$, then the diamagnetic contribution for ThMo$_2$Si$_2$C can be estimated as $\chi_{\rm{core}} = -1.01\times10^{-4}$ emu$\cdot$mol$^{-1}$ using the data listed by G. A. Bain and J. F. Berry.\cite{Pascal} Thus the Pauli-paramagnetic contribution reaches $\chi_{\rm{P}} = 2.3\times10^{-4}$ emu$\cdot$mol$^{-1}$. Then, the Wilson ratio is determined as $R_{\rm W}$ = 1.48,  which should be close to 1 in a noninteracting Fermi-liquid system. This indicates that the compound contains considerable electron--electron interaction.

Another important parameter characterizing Fermi-liquid properties is the Kadowaki--Woods ratio ($R_{\rm{KW}} = A/\gamma^2$), which presents the relationship between the electron--electron scattering rate and the electron-mass renormalization.\cite{KW-TM, KW-HF, KW-oxide, KW-normalized} This ratio is found to be approximately constant within a certain class of materials. For transition metals and heavy-fermion compounds, $R_{\rm{KW}}$s are typically 0.4 and 10 $\mu\Omega\cdot$cm$\cdot$mol$^2\cdot$K$^2\cdot$J$^{-2}$, respectively. Interestingly, our results provide a much larger $R_{\rm{KW}}$ = 212 $\mu\Omega\cdot$cm$\cdot$mol$^2\cdot$K$^2\cdot$J$^{-2}$ for ThMo$_2$Si$_2$C. Note that $R_{\rm{KW}}$ is sometime material-dependent and sensitive to dimensionality, electron density, multiband effect, and the interlayer hopping integral.\cite{KW-oxide,KW-normalized,NaCoO} To our knowledge, large $R_{\rm{KW}}$ values are usually observed in strongly correlated transition-metal oxides or organic superconductors, such as La$_{1.7}$Sr$_{0.3}$CuO$_4$ (52 $\mu\Omega\cdot$cm$\cdot$mol$^2\cdot$K$^2\cdot$J$^{-2}$), Na$_{0.7}$CoO$_2$ (600 $\mu\Omega\cdot$cm$\cdot$mol$^2\cdot$K$^2\cdot$J$^{-2}$), and $\beta$-ET$_2$I$_3$ (347 $\mu\Omega\cdot$cm$\cdot$mol$^2\cdot$K$^2\cdot$J$^{-2}$).\cite{KW-oxide, LaSrCuO, NaCoO, organic-1, organic-2} Aside from the intrinsic mechanism for large $R_{\rm{KW}}$s, we note that the magnetic impurities in a polycrystalline sample may affect the fitting parameter $A$ in the $\rho(T)$ equation, which in turn can lead to overestimation of $R_{\rm{KW}}$. Thus, high-quality samples, preferably single crystals, are needed to clarify this issue.

Now let us discuss the electronic states in the CeCr$_2$Si$_2$C-type family. So far, superconductivity has only been observed in members containing Ti$_2$O or Mo$_2$C planes. For Ti$_2$O-based compounds, superconductivity competes with the DW ordering state, which is ascribed to Fermi-surface nesting or $p$--$d$ orbital ordering.\cite{BaTi2Sb2O-FDL, BaTi2As2O-CXH} For compounds containing Mo$_2$C-plane, the DW ordering is absent. We note that the DW state is also absent among the members containing Cr$_2$C-plane. This suggests the origin of superconductivity in the carbides is different from that in the Ti$_2$O-based compounds. Very recently, neutron diffraction works on the nonsuperconductive compound UCr$_2$Si$_2$C identified an antiferromagnetic ordering of Cr magnetic moment ($\sim$0.6 $\mu_{\rm B}$) at 300 K.\cite{UCr2Si2C} However, similar experiments on the isostructural compounds $Ln$Cr$_2$Si$_2$C ($Ln$ for lanthanides) did not recognize any Cr moment.\cite{RCr2Si2C-PRB} For the compounds based on Mo$_2$Si$_2$C layers, superconductivity was only observed in ThMo$_2$Si$_2$C rather than $Ln$Mo$_2$Si$_2$C.\cite{CeMo2Si2C,PrMo2Si2C} Considering the different nominal valence of (U/Th)$^{4+}$ and $Ln^{3+}$, one may conclude that the electron configuration plays a crucial role in determining the $d-$orbital states in Cr$_2$C and Mo$_2$C planes. As the chemical valence of Cr/Mo is also affected by the interlayer Si--Si bond order, which is sensitive to the Si--Si bond length, additional theoretical and experimental works are needed to determine the configuration of the $d-$orbital in the materials.

\section{\label{sec:level4}Concluding Remarks}

We have synthesized a new quaternary compound ThMo$_2$Si$_2$C with superconducting $T_{\rm c}$ = 2.2 K. This is the first superconductor among the CeCr$_2$Si$_2$C-type carbides. The lower, upper, and thermodynamic critical fields at zero temperature are determined to be $H_{\rm{c1}}$ = 16 Oe, $H_{\rm{c2}}$ = 2660 Oe and $H_{\rm c}$ = 183 Oe, respectively. Then, the penetration depth, coherent length, and Ginzburg--Landau parameter are calculated accordingly as $\lambda$ = 5280 \AA, $\xi$ = 352 \AA, and $\kappa$ = 15. Although the jump of specific heat due to the superconducting transition deviates from that predicted by the BCS theory, the superconducting energy gap derived by fitting the specific heat below the $T_{\rm c}$ precisely matches the theoretical value. The Wilson ratio and Kadowaki--Woods ratio are obtained as $R_{\rm W}$ = 1.48 and $R_{\rm KW}$ = 212. Our data indicate that ThMo$_2$Si$_2$C is a BCS superconductor with considerable electron--electron interaction. More efforts should be made to explore new superconductors among compounds with similar crystal structures.

\begin{acknowledgments}
This work was supported by the National Key Research and Development Program of China (Grant No. 2017YFA0303002) and the Natural Science Foundation of Shandong Province (Grants No. ZR2019MA036 and ZR2016AQ08).
\end{acknowledgments}

\end{document}